\documentclass{article}

\usepackage{arxiv}
\usepackage{graphicx}
\graphicspath{ {./fig/} }

\usepackage[utf8]{inputenc} % allow utf-8 input
\usepackage[T1]{fontenc}    % use 8-bit T1 fonts
\usepackage{hyperref}       % hyperlinks
\usepackage{url}            % simple URL typesetting
\usepackage{booktabs}       % professional-quality tables
\usepackage{amsfonts}       % blackboard math symbols
\usepackage{nicefrac}       % compact symbols for 1/2, etc.
\usepackage{microtype}      % microtypography
\usepackage{lipsum}

\title{A novel application of solid state detectors for high precision, low systematics measurement of beta decay energy spectra of interest for neutrino and nuclear physics.}

\author{
M.~Biassoni \\
  Istituto Nazionale di Fisica Nucleare\\
  \And
  M.~Carminati \\
  Politecnico di Milano\\
  \And
  L.~Coraggio \\
  Istituto Nazionale di Fisica Nucleare\\
  \And
  O.~Cremonesi \\
  Istituto Nazionale di Fisica Nucleare\\
  \And
  C.~Fiorini \\
  Politecnico di Milano\\
  \And
  C.~Gotti \\
  Istituto Nazionale di Fisica Nucleare\\
  \And
  M.~Gugiatti \\
  Politecnico di Milano\\
  \And
  L.~Pagnanini \\
  Università degli Studi di Milano Bicocca\\
  \And
  M.~Pavan \\
  Università degli Studi di Milano Bicocca\\
  \And
  S.~Pozzi \\
  Università degli Studi di Milano Bicocca\\}

\begin{document}
\maketitle

\begin{abstract}
This project is focused on the development of a novel strategy for the precise measurement of beta decay energy spectra. The exact determination of beta spectra has wide implications in the particle and nuclear physics fields: from sterile neutrino searches \cite{Adhikari 2016} to validation of nuclear models of interest for double beta decay searches \cite{Suhonen 2017}, to reactor neutrino experiments.
The experimental strategy is focused on the mitigation of the systematic uncertainty in the determination of the spectral shape related to the energy response of the detector.
The plan is to use Silicon Drift Detectors (SDDs), exploiting their excellent energy resolution and response uniformity, the thin dead layer and the fast signal that allows high rate operations.
A complete modeling of SDDs response to the interaction of beta particles via numerical simulations, validated with dedicated measurements, aims at identifying and canceling the detector-related systematics. Similar attention will be devoted to the integration of beta radioactive sources that preserve the energy information carried by the emitted electrons, and to a veto system with a full solid angle coverage to intercept any escaping particle potentially carrying a fraction of the original electron energy.
\end{abstract}

% keywords can be removed
\keywords{Nuclear Physics \and Spectroscopic and Spectrometric Techniques \and Physics of Detectors \and Elementary Particles \and Beta Decay}

\section{Introduction}
The focus of the project is the development of a novel detection strategy to perform a high precision measurement of the energy spectra of beta decays of interest for the physics community.
The aim is to design, build and operate an innovative and flexible spectrometer based on Silicon Drift Detectors (SDDs) to perform high resolution beta spectroscopy of isotopes with transition energies ranging from $\sim$10~keV to $\sim$1~MeV.\\
The project will demonstrate that SDDs are flexible devices suitable for high resolution beta spectroscopy of a variety of isotopes in large counting rate applications, as long as systematic uncertainties related to the detector response to the interaction are under control: energy scale, linearity of the energy response, possible deformations of the reconstructed energy spectrum due to dead layers, incomplete formation or collection of the signal, geometrical efficiency of the source-detector combination and partial containment of the electron kinetic energy due to secondary radiation escaping the detector.
Cryogenic calorimeters \cite{Twerenbold 1996}, with similar energy resolution but a simpler- and easier-to-model detector response (and complementary systematic uncertainties), will be used for a comparative study of the spectrometer response function.\\
To isolate and study the systematic contributions to the uncertainty, the statistical error on the measured spectra has to be reduced to a negligible level. This requires careful consideration of the choice of the source activity, measurement campaigns duration and environmental background.\\
The modeling of the detector with Monte Carlo and numerical simulations is a key element of the project, as well as its validation with dedicated measurements aimed at characterizing the detector performance in different operating conditions (energy range, interaction rate, incidence angle and position). These measurements can be performed both with electrons from radioactive and artificial sources (e.g. an electron microscope), and with X-ray and low energy gamma sources, in order to isolate and study the different aspects of the detector response.\\
The development of the experimental setup, as well as the definition of procedures and protocols for its operation, are as important as the complete understanding of the SDD response. This guarantees the reproducibility of the measurement and simplifies the comparison of results for different isotopes, once more aiming at the reduction of systematic uncertainties. The setup will include a main detection module where the SDD will be cooled to an optimal working temperature and connected to low-noise electronics to enhance performance, as well as a veto system to reject events characterized by an incomplete energy deposition in the SDD.\\
The  beta-emitting isotopes candidate for the measurement  are selected based on the scientific impact of the reconstruction of the corresponding energy spectrum. Particular focus will be put on multiple-forbidden unique and non-unique beta decays of heavy isotopes, as well as isotopes responsible for the emission of neutrinos studied by reactor neutrino experiments.\\
The precise reconstruction of energy spectra of multiple-forbidden, unique and non-unique beta decays would have a notable impact in the field of nuclear physics, leading to a refinement of existing nuclear models and to the ruling out of those unable to reproduce experimental observations.\\
In particular, the aim is to contribute to the solution of the puzzle related to the need for quenching the axial coupling constant g$_A$ in nuclear many-body calculations to reproduce beta-decay data \cite{Suhonen 2017}. As a matter of fact, improving the knowledge of the quenching mechanism in the nuclear theory translates into a beneficial effect for the double beta decay searches as well \cite{Iachello 2015}. 
This field would be strongly influenced by these results, as both g$_A$ and the calculation of nuclear matrix elements play a critical role in estimating the sensitivity of an experimental technique to new physics. 
Furthermore, the improved precision and reliability of nuclear models would result in a more accurate prediction of the shape of neutrino energy spectra. Nuclear reactor neutrino experiments, as well as geo-neutrino measurements, would benefit from the resulting mitigation of an important source of systematic uncertainties affecting their results. \\
Finally, the study of tritium beta spectrum with magnetic spectrometers coupled to solid state detectors \cite{Adhikari 2016}, searching for deformations ascribable to the existence of a keV-scale sterile neutrino, requires an extremely precise knowledge of the instrumental effects, that will be studied in this project, that could result in a spectral distortion.

\section{Scientific goal}
\label{sec:scientific goal}
The main targets of the project are:
\begin{itemize}
\item develop and optimize a spectrometer based on Silicon Drift Detectors (SDDs) and a corresponding measurement protocol (source production, background reduction, data analysis) for the reconstruction of beta decay energy spectra with end point in the 10 keV - 1 MeV range, with reduced systematic uncertainty on the spectral shape over the full spectrum extension;
\item optimize the source production and source-detector interaction to further mitigate the instrumental effects contributing to the systematic uncertainty;
\item perform high precision measurement of the energy spectrum of multiple beta emitting isotopes, selected based on their relevance for the nuclear and particle physics community (nuclear models, sterile neutrino searches, double beta decay, reactor neutrino measurements, geoneutrinos).
\end{itemize}

\section{State of the art}
\label{sec:state of the art}
The emission of negatively charged particles from a nucleus is one of the first discovered forms of radioactivity. Beta decays have contributed enormously to the development of nuclear physics and to our understanding of weak interactions and neutrinos. Despite a strong effort to find a satisfactory description of this form of radioactivity, the theory describing nuclear beta decay is very complicated and still incomplete. 
Direct measurements which could provide important information for its development are very difficult and, with few exceptions, have been practically abandoned in the past 60 years \cite{Gove 1971, Behrens 1976, Daniel 1966, Paul 1966}.
In the past decade, however, interest in beta decays has flourished and they are now the keystone of a number of academic studies and practical applications.
When coupled to the constant development of technology and theory, new perspectives are possible.
Indeed, so far beta decays have been studied using gas counters or scintillators which are characterized by limited energy resolution and efficiency and can provide therefore only inaccurate measurements. 
Systematic surveys of beta emitters are available in the literature but they are often limited to measurements of the decay half lives carried out in the 1960s. Recent attention to beta processes of interest in astrophysics has triggered the measurement of heavy short-lived isotopes, however limited again to an estimate of the half life.

The combination of new high-resolution technologies for the measurement of the electron spectrum, coupled to massive conventional detectors to gather all the available information (escaping electrons or photons) can provide measurements of unprecedented precision of the full spectrum and of the lifetime of medium and long lived isotopes which could help solving some of the open questions in fundamental physics and have a direct impact on practical applications.

The spectroscopy of electrons in the 10 keV - 1 MeV energy range is challenging, when compared to other radiation types, due to the typically large systematic effects related to the detector response to the electrons interaction. These effects depend on many different aspects of the detector technology:
\begin{itemize}
    \item dead layer: a fraction of the volume, typically close to the surface where the electrons enter the detector, where no (or strongly non-linear) signal is generated in response to an energy deposition;
    \item response non-linearity: the signal generated doesn’t depend linearly on the amount of deposited energy, and the non-linearity cannot be easily removed with simple calibration procedures;
    \item back-scattering: after some interactions in the detector material the electron is reflected out of the detector, and only a fraction of its original energy is deposited and recorded in the spectrum;
    \item X-ray and bremsstrahlung escape: low energy photons are created as a result of ionization and electron scattering in the nuclear field. The corresponding energy is lost if the photons escape the detector active volume;
    \item containment efficiency: depending on the size and geometry of the detector, an electron could escape the active volume before depositing all its energy;
    \item position dependence: the signal generated for a given energy deposition could depend on the position where the energy was deposited;
    \item pile-up: some detectors, depending on the signal speed and the interaction rate, are subject to wrong energy reconstruction if two interactions are too close in time, and the corresponding signals overlap without being identified and rejected.
\end{itemize}

As all these effects can lead to a distortion of the reconstructed energy spectrum, their understanding and characterization is mandatory in order to improve the precision of the proposed SDD-based spectrometer. 

\section{Spectrometer structure}
Thanks to their low output capacitance, SDDs are high resolution, fast response semiconductor detectors. Their application to high precision beta spectroscopy is a promising and largely unexplored field, that would profit from the detailed study and mitigation of the above-mentioned sources of systematic effects.
SDDs have been already successfully employed for X-ray spectroscopy measurements either in scientific research \cite{Bazzi 2013} as well as in commercial applications, e.g. as X-ray detector in SEM (Scanning Electron Microscopes) microanalyzers. The introduction of monolithic arrays of SDDs \cite{Fiorini 2010, Bufon 2017} has allowed to cover large sensitive areas with low dead margins and develop high-throughput spectrometers. Despite the success of SDDs for X-ray detection, its use for electrons detection is a rather unexplored field \cite{Cox 2011} and this represents one of the tasks of the present proposal. 
We propose to use the SDD technology available at the Fondazione Bruno Kessler (FBK) of Trento (Italy), where SDDs have been produced successfully both for direct X-ray detection \cite{Quaglia 2015, Bertuccio 2016} and for scintillation light readout \cite{Fiorini 2013}. 

The entrance window of the SDD represents one peculiar technological aspect in this project, as electrons unavoidably deposit energy in the non-active layer of the detector entrance window. An extremely uniform and thin entrance window, shallower than 100 nm, is needed to be sensitive to electrons energies of few keV \cite{Kanaya 1972}. Although a different mechanism rules X-ray and electron absorption in silicon, an efficient detection of electrons would not be possible if the dead layer in the entrance window were not small enough also for soft X-rays detection. Recently \cite{Borghi 2018},  measurements of soft X-rays down to C-Kalpha line (277eV) have been performed with FBK SDDs. In addition, the same entrance window has been implemented in SDDs able to detect scintillation photons at short wavelength, as low 360nm from LaBr3(Ce) scintillators \cite{Fiorini 2013}. This performance is possible only with a dead-layer in the entrance window smaller than 100nm \cite{Lechner 1995, Hartmann 1996, Eggert 2005}.
In the first year of this project we'll characterize the SDDs response to electrons using existing prototypes. In the following years we'll produce SDDs specifically designed to optimize the response to electrons and in a format suitable for a demonstrator for beta spectra measurements.
The high performance of SDDs can be fully exploited by the use of low-noise CMOS charge preamplifiers \cite{Bombelli 2011},  assembled close to the SDD anode. The use of processing ASICs \cite{Schembari 2017} after the preamplifier allows possible scaling of a SDD-based X-ray detector to arrays of hundreds of units as, for example, in the case of the upgrade of the SIDDHARTA experiment to 200cm2 of SDDs \cite{Butt 2016}. 
New integrated readout electronics, build to operate the SDDs with high resolution and linearity in the extended energy range of beta spectra, will be developed specifically for this project. Timing features of the new electronics will allow synchronization and reconstruction of signals among different SDDs and other fast detectors (scintillators).

The spectrometer structure will be conceived to integrate multiple detectors (see Fig.1 for a conceptual sketch), and the ancillary elements required for their operation, in a design optimized to maximally exploit the performance of the SDDs while minimizing the effect of residual uncertainties in their response.

\begin{figure}
  \centering
    \includegraphics[scale=0.18]{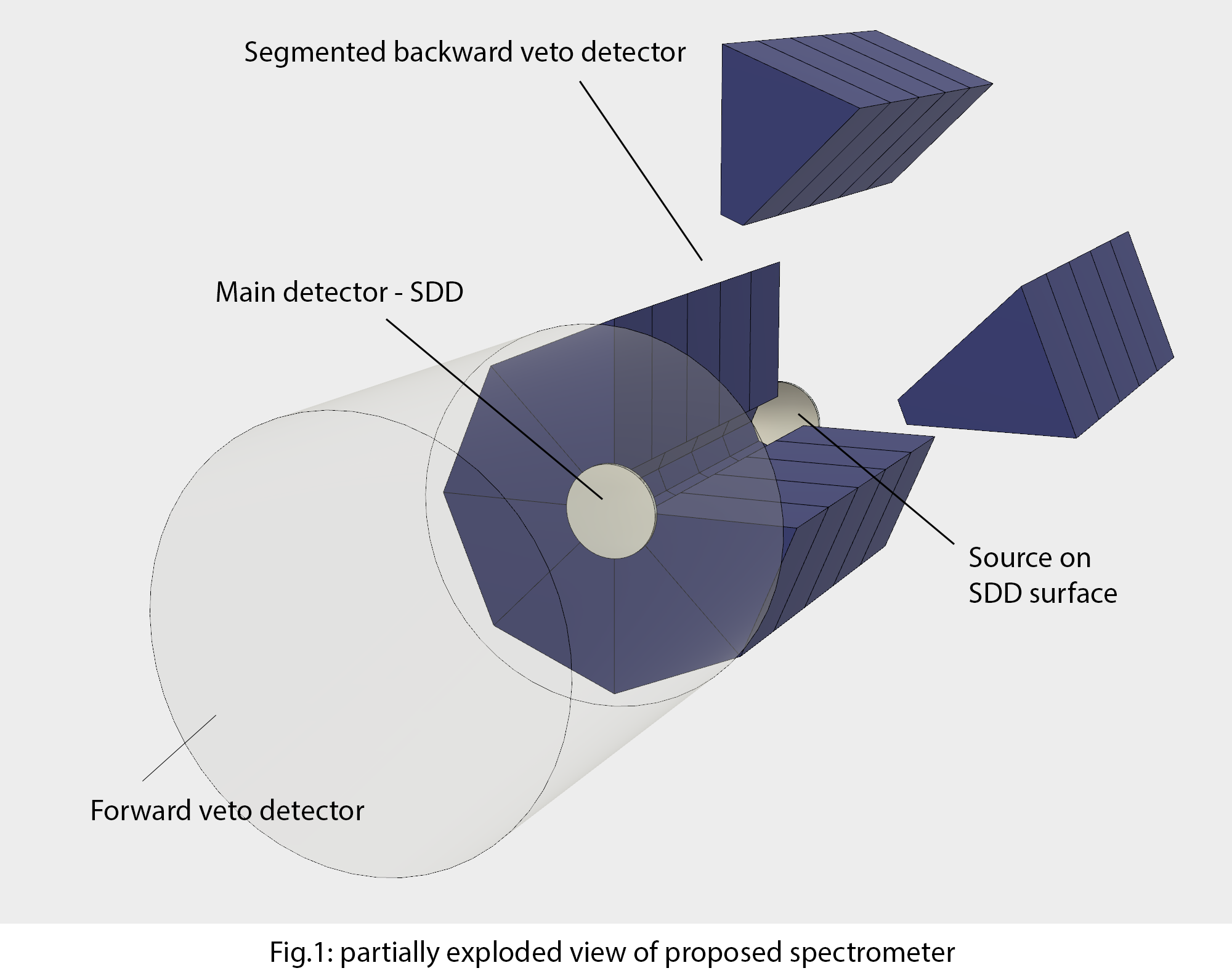}
  \label{fig:sketch}
\end{figure}

The core of the system will be an SDD where the electron coming from the beta decaying source must be contained. Given the technological constraints on the SDD thickness, a stack of two (or in principle more) SDDs may be required to fully stop the electron, depending on the transition energy (Q-value) of the decay under study. This SDDs stack will be called the main detector in the following.
To clearly separate signal-like events from spurious events based on their topology, ancillary detectors are required. In order to correctly reconstruct their energy, in fact, electrons must be completely contained in the main detector, while events where a fraction of the energy is deposited outside of it must be rejected (see Fig.2 for examples of fully contained and vetoed event topologies, and Fig.3 for an example of the effect of the proposed veto on the reconstructed spectrum of 500 keV electrons). These events are mainly due to:
\begin{itemize}
    \item electrons that deposit only a fraction of their energy in the main detector before escaping its active volume (due to back-scattering or non-unitary geometrical efficiency of the stack);
    \item Si X-rays produced as a by-product of the primary ionization process that escape from the SDD;
    \item bremsstrahlung photons produced by the deceleration of electrons interacting with the detector material and escaping from the SDD.
\end{itemize}

\begin{figure}
  \centering
    \includegraphics[scale=0.135]{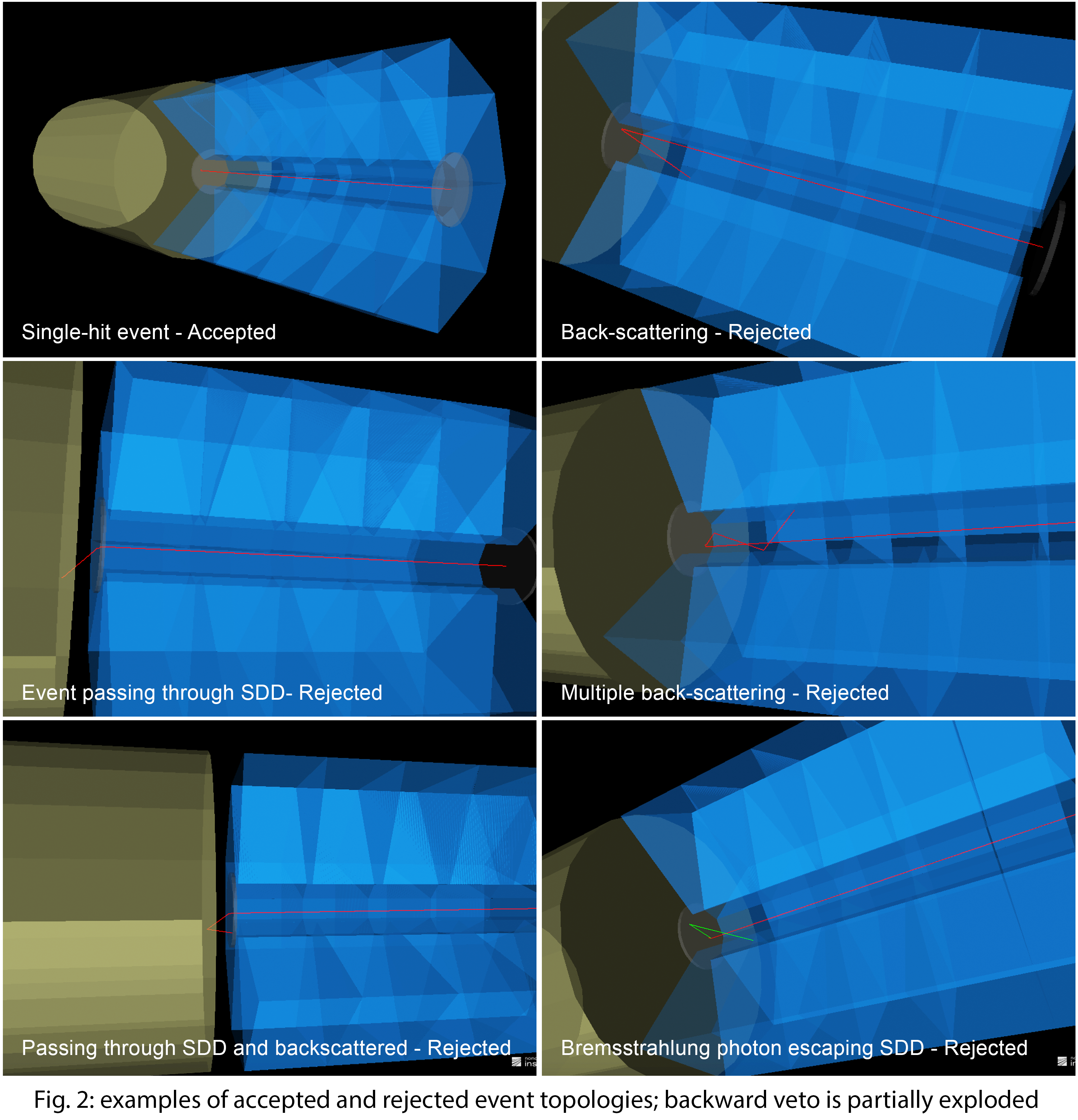}
  \label{fig:topology}
\end{figure}

\begin{figure}
  \centering
    \includegraphics[scale=0.75]{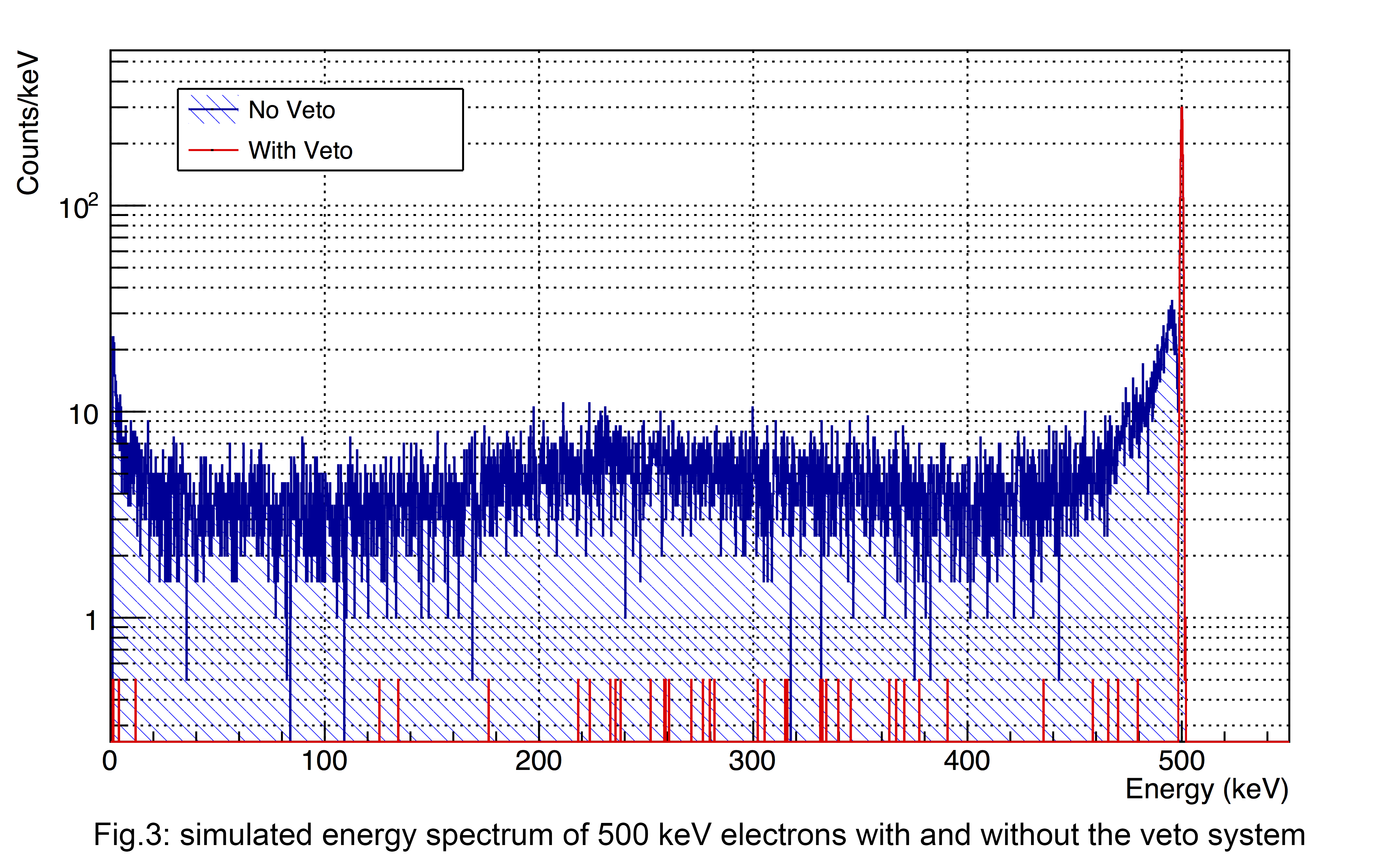}
  \label{fig:spectrum}
\end{figure}

The veto detectors must have high detection efficiency for photons in the few hundreds of keV range and very low threshold (down to a few keV or less) for both photons and electrons and provide large angular coverage (requiring minimal passive materials between the SDDs and the veto detectors). 
Fast inorganic scintillators, arranged in closely packed arrays to maximize the angular coverage and read by single-photon sensitive photo-detectors, are the baseline choice.

The exact geometry and size of the system (SDDs, veto detectors, source and passive elements) and the minimum performance of the different detectors will be optimized based on the output of a Monte Carlo-based study.

Depending on the outcome of the study of the SDD response as a function of the electron incidence angle, the necessity of collimating the source will be evaluated, especially for lower energy sources where the effect of the device dead layer is expected to have a larger relative effect. The consequences on the source design will be discussed in the following section.

The spectrometer (main detector and veto detectors) will be contained in a custom-designed vacuum chamber, also containing the cooling system used to keep the detectors at the operating temperature. Front-end CMOS preamplifiers will be located inside the vacuum chamber  and assembled in close vicinity to the SDD. The remaining part of the processing electronics, namely analog ASICs, will be placed immediately outside the vacuum chamber.
New ASICs will allow to cover an extended dynamic range with good resolution and linearity and will include the possibility to record time-stamps of the events to allow to record coincidences between various SDDs and other detectors.

Finally, the need for an active or passive shielding against environmental radioactivity will be evaluated, based on simulations and depending on the source activity that the system will be able to sustain.

\section{Beta source}\label{sec:source}
In order to reduce the range of incident angles of the electrons on the main detector (thus minimizing the effect of the dead layer), the source must be deposited on an element different from the main detector (i.e. the SDD where the energy measurement is performed). In this case the source support will need to be an active detector as well (another SDD), in order to veto the events where the primary electron is emitted in the wrong direction but later back-scattered in the spectrometer acceptance angle, thus reaching the main detector with degraded energy. 

The source will therefore be placed, by chemical or electro-chemical deposition procedures or exposure to low-energy radioactive ions beams, on the surface of an SDD. The possible effects of the deposition on the performance of the detector, and consequently the optimized deposition parameters, will be studied with device simulations and real data acquired with single SDD prototypes. This approach for the detector-source coupling is considered safe and relatively easy to implement.

It is worth noting that a relatively wide choice of isotopes to be used during the project development are commercially available or easy to procure, and the overall activity of the sources is expected to be moderate to negligible from a radio-protection perspective.

\section{Methodology applied to detector characterization}\label{sec:methodology}
The study of the SDDs energy response will be performed by comparing numerical and Monte Carlo simulations of the detectors and their setup with the data collected during measurement campaigns where the SDDs are exposed to selected radiation sources.

In order to explore different aspects of the detector response, these sources will be employed:
\begin{itemize}
    \item X-rays in the 1 keV - 10 keV energy range, to generate single electrons by photo-electric effect in the volume of the SDD. X-rays interactions are relatively simple to simulate, but their interaction in the surrounding environment and in the source itself must also be taken into account. The SDDs can be illuminated uniformly or locally by using collimators. Information about the uniformity of the detector response can be extracted from a comparison between the data and numerical simulations;
    \item electrons from a SEM (Scanning Electron Microscope): intrinsically collimated and mono-energetic, this source can be used to study the detector response as a function of electron energy in the 5 keV - 30 keV range, interaction position and incident angle with good resolution in all these variables. A set of measurements with different incident angles will be used to estimate the thickness of the SDDs entrance window dead layer. In order to perform such tests, a setup containing the SDD prototype, the front-end preamplifier and supporting hybrid integrated circuit will be realized to be compatible with operation inside the SEM sample chamber;
    \item beta sources: one or more beta sources with simple decay schemes must be identified to be used as reference. SDDs will be exposed to these sources to perform a direct validation of the detector response model, by comparing the experimentally determined spectrum with the theoretical expectation.
\end{itemize}

The scintillation detectors used for the veto system must be carefully characterized in terms of energy threshold. Their capability of detecting with high efficiency small quantities of energy escaping (in the form of back-scattered electrons, X and gamma rays) from the main detector is essential to identify (and reject) beta events whose original energy cannot be precisely determined. This characterization will be performed by detecting the X-ray fluorescence of low-Z elements (with sub-keV Ka X-rays).

Finally, aiming at a robust and model-independent confirmation of the capability of the full spectrometer system to provide a precise determination of the energy spectrum, a comparative measurement with a different detector technology of one reference beta spectrum is beneficial. Cryogenic calorimeters of large mass are a good option, thanks to a very simple-to-model detector response (no dead layer, high linearity, no position dependence, very high containment efficiency of photons). The sources of spectral shape-deforming systematic effects are very different from those characteristic of the SDD based spectrometer under study, being mainly due to the slow detector response and sensitivity to environmental background. 

\section{Conclusions}\label{sec:conclusions}
In this work we present a novel application of SDD detectors to the measurement of nuclear beta decays. A comprehensive study of the experimental setup, based on SDDs for electron spectroscopy and scintillating crystals for active background rejection, SDD response function characterization procedures and beta source preparation has been performed, resulting in a proposal whose goal is to minimize any systematic effect that could affect the experiment capability of characterizing the shape of the beta spectrum in a wide energy range.
This paper is based on a project submitted for founding to MIUR - PRIN 2017 announcement.

\bibliographystyle{unsrt}  

%%% Comment out this section when your \bibliography{references} is enabled.

\end{document}